\pdfoutput=1
\documentclass[a4paper,american,citeautoscript,floatfix,pdftex,superscriptaddress,twoside,%
aps,prl,%
longbibliography,%
reprint,twocolumn,
]{revtex4-1}%
\usepackage{amsfonts,amsmath,amssymb}
\usepackage{commath}
\usepackage[T1]{fontenc}
\usepackage{graphicx}%
\usepackage[utf8]{inputenc}
\usepackage{microtype}
\usepackage[obeyFinal,textsize=footnotesize]{todonotes}
\usepackage{xspace}
\usepackage{hyperref, hypernat}
\usepackage[todos]{CMImacros}

\newcommand{\cfeldesy}{\affiliation{Center for Free-Electron Laser Science, Deutsches
      Elektronen-Synchrotron DESY, Notkestraße 85, 22607 Hamburg, Germany}}%
\newcommand{\uhhcui}{\affiliation{The Hamburg Center for Ultrafast Imaging, Universität Hamburg,
      Luruper Chaussee 149, 22761 Hamburg, Germany}}%
\newcommand{\uhhphys}{\affiliation{Department of Physics, Universität Hamburg, Luruper Chaussee 149,
      22761 Hamburg, Germany}}%
\newcommand{\ucl}{\affiliation{Department of Physics and Astronomy, University College London, Gower
      Street, WC1E 6BT London, United Kingdom}}%
\newcommand{\ayemail}{\email[]{andrey.yachmenev@cfel.de}}%
\newcommand{\cmiweb}{\homepage{https://www.controlled-molecule-imaging.org}}%

\begin{document}
\title{Climbing the rotational ladder to chirality}
\author{Alec Owens}\cfeldesy\uhhcui
\author{Andrey Yachmenev}\ayemail{}\cmiweb\cfeldesy\uhhcui
\author{Sergei N. Yurchenko}\ucl
\author{Jochen Küpper}\cfeldesy\uhhcui\uhhphys
\date{\today}
\begin{abstract}\noindent%
   Molecular chirality is conventionally understood as space-inversion-symmetry breaking in the
   equilibrium structure of molecules. Less well known is that achiral molecules can be made chiral
   through extreme rotational excitation. Here, we theoretically demonstrate a clear strategy for
   generating rotationally-induced chirality (RIC): An optical centrifuge rotationally excites the
   phosphine molecule (PH$_3$) into chiral cluster states that correspond to clockwise
   (\emph{R}-enantiomer) or anticlockwise (\emph{L}-enantiomer) rotation about axes almost
   coinciding with single P--H bonds. Application of a strong dc electric field during the
   centrifuge pulse favors the production of one rotating enantiomeric form over the other, creating
   dynamically chiral molecules with \emph{permanently} oriented rotational angular momentum. This
   essential step toward characterizing RIC promises a fresh perspective on chirality as a
   fundamental aspect of nature.
\end{abstract}
\maketitle

Chiral molecules exist in two structural forms known as enantiomers, which are the mirror image of
each other and thus non-superimposable by translation and rotation. The wavefunctions of the
enantiomers have no definite parity, \ie, space-inversion symmetry, as they tunnel periodically from
one structure to the other on a timescale determined by the energy barrier separating the two
enantiomeric forms. Molecular chirality, in the traditional sense, is associated with the barriers
formed by the static potential energy surface, typically so large that the enantiomers do not
readily interconvert at room temperature and can be separated by chemical means and stored. There
is, however, the possibility to induce and control chirality in achiral molecules through extreme
rotational excitation into states where centrifugal distortions produce equivalent asymmetric
structures separated by high kinetic energy barriers~\cite{Bunker:JMolSpec228:640}. The appearance
of such chiral high angular momentum states is closely linked to the energetic clustering of
rotational states, an effect displayed by certain polyatomic molecules exhibiting local mode
vibrations~\cite{Jensen:WCMS2:494}. Given the importance of chirality to our understanding of
molecular and material behavior, the ability to create chirality in achiral systems is of great
fundamental interest. For example, it has recently been shown that isotopic substitution can induce
chirality in achiral molecules and catalyze asymmetric reactions~\cite{Kawasaki:Science324:492,
   Matsumoto:ACIE55:15246}.

Very little is known about the phenomenon of rotationally-induced chirality (RIC), but
ultrashort-pulse strong-field laser physics has brought new impetus to the preparation of molecules
in highly excited rotational states. The main obstacle to overcome when generating large amounts of
rotational excitation is that selection rules permit only small changes in the overall angular
momentum $J$. With the development of clever adiabatic and non-adiabatic
approaches~\cite{Ohshima:IRPC29:619, Milner:ACP159:395}, this is now far less challenging and it is
possible to efficiently create rotational wavepackets with a narrow and well-defined distribution of
states. Molecular superrotors, as they are known in the literature, display novel behavior. For
example, they are far more resistant to collisions and reorientation~\cite{Yuan:PNAS108:6872,
   Milner:PRL113:043005, Khodorkovsky:NatComm6:7791} making them interesting objects for
scattering~\cite{Mullin:JPCA119:12471}, spectroscopy~\cite{Yuan:FD150:101}, and
dynamics~\cite{Milner:PRX5:031041}. Superrotors provide an ideal testing ground for exploring RIC,
and by learning to control and characterize this dynamic effect, \eg, the optical activity, there is
an opportunity to gain new insights into chirality, specifically motion-dependent chiral systems and
the nature of their interactions.

\begin{figure*}
   \centering%
   \includegraphics[width=0.75\linewidth]{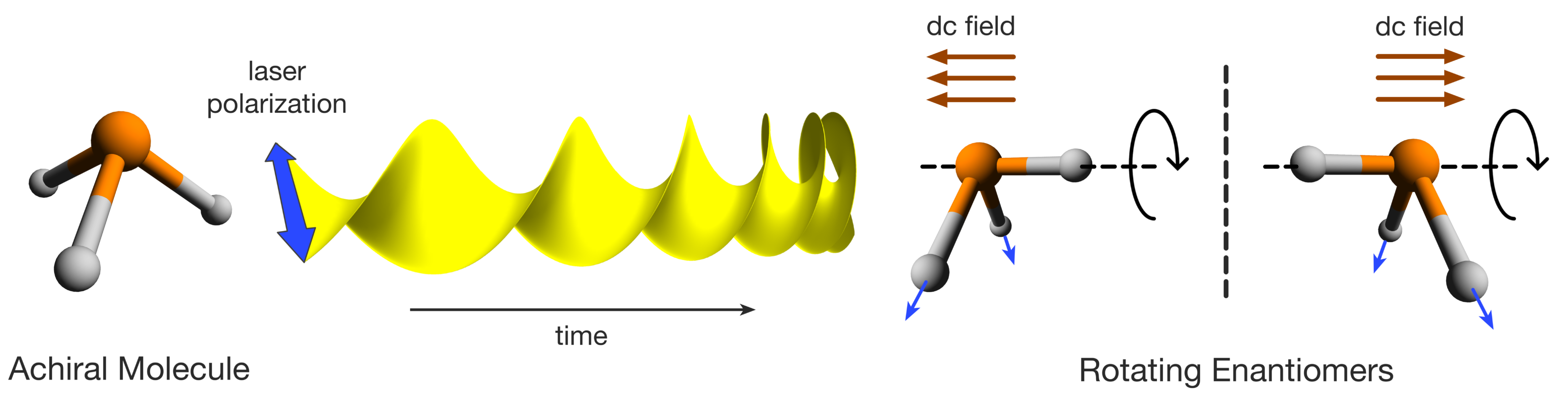}
   \caption{Rotationally-induced chirality in PH$_3$ using an optical centrifuge in conjunction with
      a static electric field. Reversing the direction of the dc field produces the other rotating
      enantiomer.}
   \label{fig:ric}
\end{figure*}
\begin{figure*}
   \centering%
   \includegraphics[width=\linewidth]{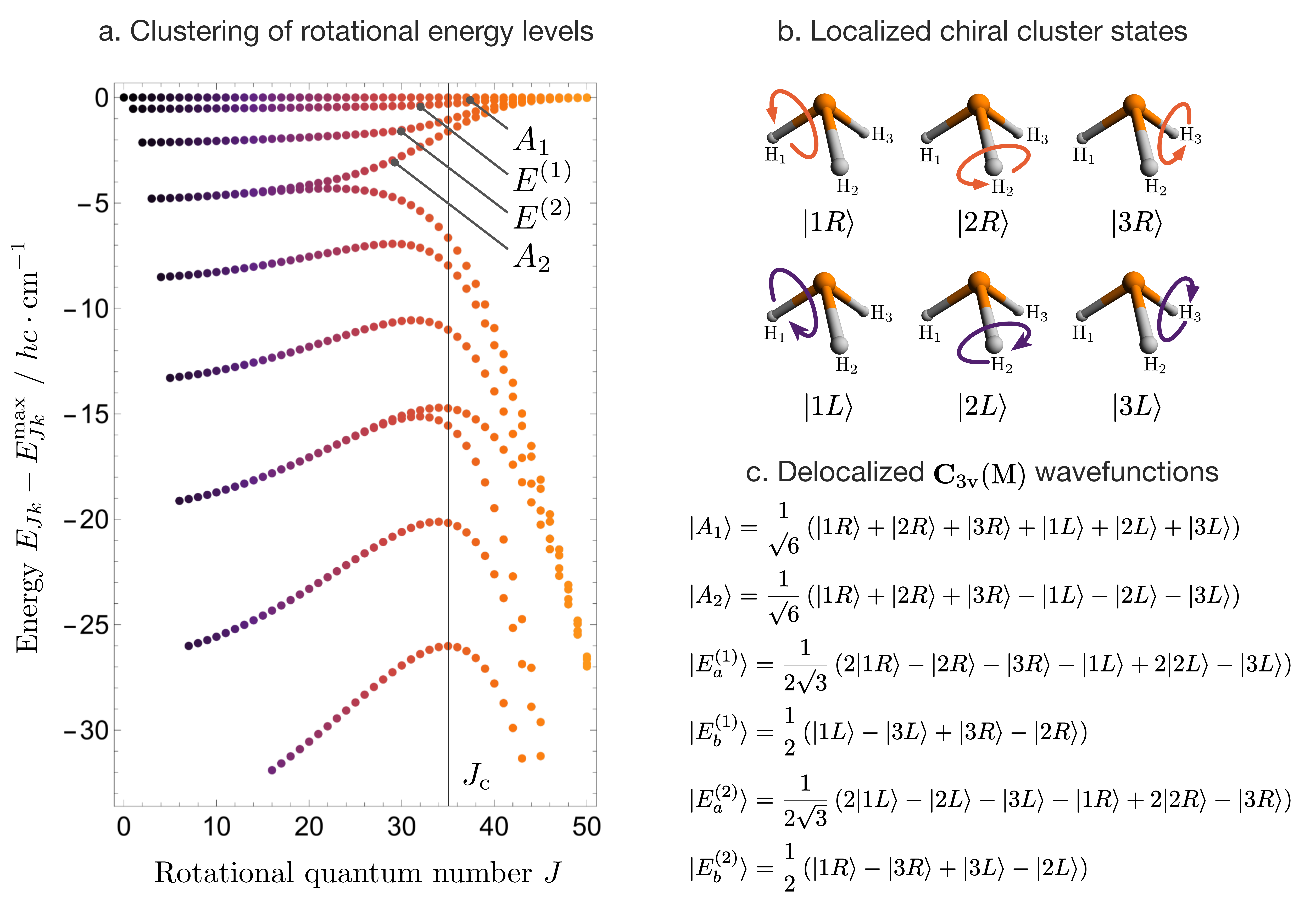}
   \caption{Rotational energy level clustering in the ground vibrational state of PH$_3$, \cf\
      \onlinecite[Fig.~2 of ref.][]{Yurchenko:PCCP7:573}. (a) The energy difference
      $E_{Jk}-E_{Jk}^{\text{max}}$ was plotted for each rotational energy level $E_{Jk}$ relative to
      the maximum energy $E_{Jk}^{\text{max}}$ in its $J$ multiplet. (b) Schematic representation of
      the localized chiral cluster states, which are separated by high kinetic energy barriers. (c)
      Wavefunctions of the delocalized rotational cluster states in the representation of the
      localized chiral cluster states.}
   \label{fig:cluster}
\end{figure*}

In this work, a clear strategy for producing dynamically chiral molecules through extreme rotational
excitation is presented. The scheme is outlined in \autoref{fig:ric} and realized through robust
quantum mechanical simulations of RIC in the phosphine molecule (PH$_3$). PH$_3$ is initially
considered in its ground rovibronic state as a symmetric top molecule possessing
$\mathbf{C}_{3v}$(M) molecular symmetry; inversion tunneling has a high barrier and is neglected. An
optical centrifuge~\cite{Karczmarek:PRL82:3420}, which is a non-resonant, linearly polarized laser
pulse that undergoes accelerated rotation about the direction of propagation, is used to
adiabatically spin phosphine into highly excited rotational states. Above the critical value of
$J_\text{c}\approx35$, the rovibrational energy levels of PH$_3$ form sixfold degenerate
clusters~\cite{Yurchenko:PCCP7:573}, which classically can be interpreted as a ``racemic mixture''
of clockwise (\textit{R}) and anticlockwise (\textit{L}) rotations about localization axes
approximately coinciding with a P--H bond, see \autoref{fig:cluster}. Due to strong centrifugal
coupling, three dynamic and symmetrically equivalent molecular structures emerge, where one of the
P--H bonds about which the molecule rotates, is shorter than the other two. The three asymmetrical
structures can rotate in either a clockwise or anticlockwise direction. They are energetically
indistinguishable from one another and separated by high kinetic energy barriers, which is analogous
to conventional ``static'' enantiomers with potential barriers. If PH$_3$ is prepared in a state
corresponding to a preferred direction of rotation in the laboratory frame, the permutation symmetry
is broken by centrifugal distortions, whilst parity is broken by unidirectional rotation. The result
is a rotating enantiomer, where clockwise rotation corresponds to the \textit{R}-enantiomer, and
anticlockwise rotation to the \textit{L}-enantiomer.

The relationship between the localized chiral cluster states and the delocalized wavefunctions in
the $\mathbf{C}_{3v}$(M) symmetry group is shown in \autoref{fig:cluster}. It is evident that any
interaction of the permanent molecular electric dipole moment with a strong electric field will mix
the delocalized wavefunctions of quasi-degenerate states, e.g., $A_1$ with $A_2$, in such a way that
combinations of states dominated by either \emph{R} or \emph{L} localized states are generated.
Hence, by applying a strong dc field during the centrifuge pulse either one of the rotating
enantiomers can be exclusively produced depending on the direction of the field. This is an
essential step toward characterizing RIC as each enantiomer can be isolated for further
investigation. Furthermore, it demonstrates that an ensemble of dynamically chiral molecules with
``permanently'' oriented rotational angular momentum can be created, where the orientation of the
angular momentum with respect to the P--H bond, about which rotation takes place, defines the
rotating enantiomer. That is, the difference between the centrifuge with and without the dc field is
the orientation of the angular momentum in the molecular frame.

It should be noted that the Raman-type excitation by the optical centrifuge couples states with the
same symmetry, whereas the dc field couples states whose product-symmetry spans the symmetry of the
dipole moment $A_2$, \ie, $A_1\leftrightarrow{}A_2$ and $E\leftrightarrow{}E$. Therefore, depending
on the symmetry of the initial rotational state, only a pair of $A_1$ and $A_2$ or a pair of
$E^{(1)}$ and $E^{(2)}$ delocalized components of the rotational cluster will give rise to the
effect of RIC.

The theoretical approach is described in detail in the supplemental material~\cite{SM:PRL:PH3:2018}.
Numerical simulations were fully quantum mechanical and considered all major ground-state electronic
correlation within the Born-Oppenheimer approximation, nuclear motion, and external field effects to
an accuracy capable of supporting high-resolution spectroscopic measurements, \ie, sub-\invcm
accuracy for vibrational and $10^{-4}$--$10^{-5}~\invcm$ for rotational energies. The stationary
rovibrational energies and eigenfunctions of PH$_3$ up to $J=60$ were computed by direct numerical
variational calculations~\cite{Yurchenko:JMS245:126, Yachmenev:JCP143:014105, Yurchenko:JCTC13:4368}
from an empirically refined, six-dimensional potential energy
surface~\cite{Sousa-Silva:MNRAS446:2337}. To describe external field effects, the matrix elements of
the electric dipole moment and electric polarizability tensor were evaluated in the basis of
rovibrational states. In quantum dynamics simulations, which employed the computer program
RichMol~\cite{Owens:JCP148:124102}, the time-dependent wavefunction was built from a superposition
of rovibrational eigenfunctions with the time-dependent coefficients obtained by numerical solution
of the time-dependent Schrödinger equation using the split-operator method.

An optical centrifuge with maximum intensity $\epsilon_0=5.2\times10^{13}$~W/cm$^2$ and chirp rate
$\beta=(2\pi{}c)^2\cdot 25$~cm$^{-2}$ was applied along the laboratory-fixed $Z$ axis for a duration
of $t=36.8$~ps. Simulations were started in the rovibrational ground state ($J=k=m=0$),
corresponding to samples that can be produced from cold beams using the electric
deflector~\cite{Chang:IRPC34:557, Horke:ACIE53:11965}. The centrifuge field aligns the most
polarizable molecular axis through the interaction with the induced molecular dipole moment,
trapping and forcing the molecule to follow the rotating polarization. PH$_3$ is predominantly
excited via $\Delta{J}=2,\Delta{m}=-2$ rotational Raman transitions through the highest energy state
of each $J$ value. Here, the quantum numbers $k$ and $m$ correspond to the projection, in units of
$\hbar$, of $J$ onto the molecule-fixed $z$ axis and laboratory-fixed $Z$ axis, respectively. For
calculations involving a static electric field, a field strength of
$1$~MV/cm~\cite{Ishida:Nature124:129} was applied along the laboratory-fixed $Z$ axis.

The phenomenon of RIC is well illustrated with the chosen laser parameters, but the centrifuge
intensity $\epsilon_0$ is expected to cause ionization of PH$_3$. However, the same conclusions can
still be reached with lower optical and static field strengths, and when using an experimentally
realistic pulse envelope for the optical centrifuge, see the supplemental
material~\cite{SM:PRL:PH3:2018}. It should be noted that because PH$_3$ is highly toxic, we
anticipate that future experiments will be performed on other systems, most likely heavier rotors,
which require weaker optical and static field strengths, allowing for the practical realization of
this proposal.

\begin{figure*}
\centering
   \includegraphics[width=0.75\linewidth]{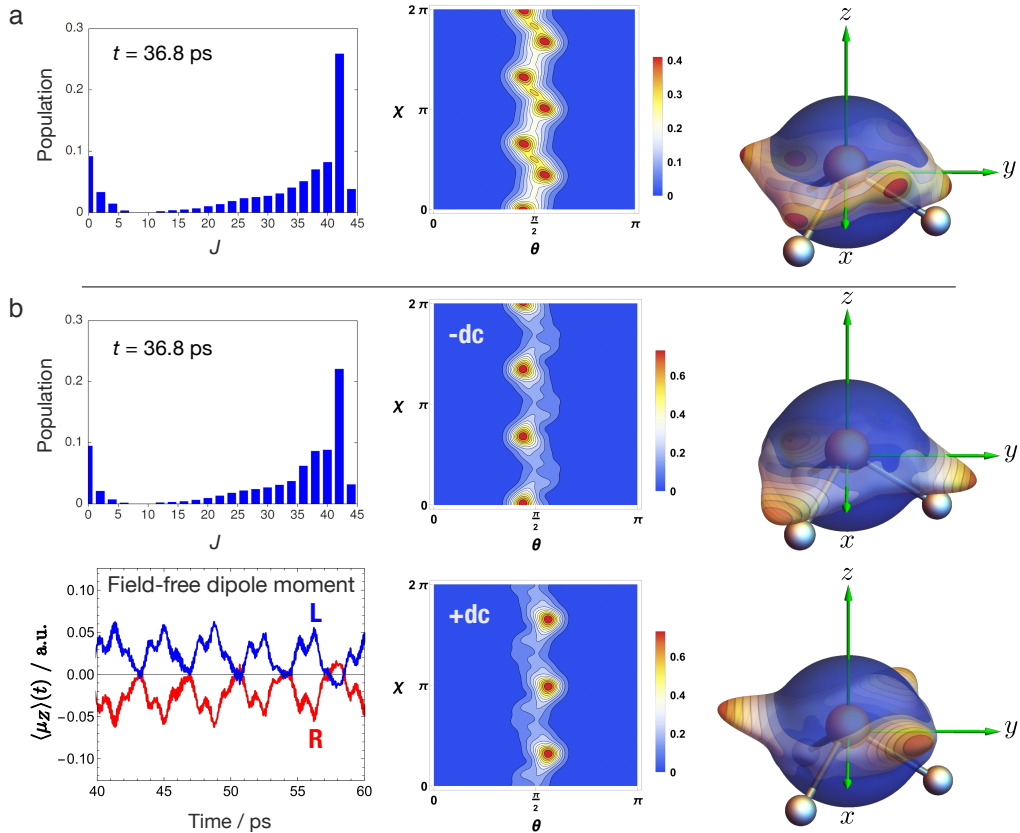}
   \caption{Wavepacket population by $m=-J$ stationary states and rotational probability density
      function $P(\theta,\chi)$, in terms of the Euler angles $\theta$ and $\chi$, at the end of the
      centrifuge pulse ($t=36.8$~ps), (a) without, and (b) with a dc field of $\pm1$~MV/cm applied.
      Wavepacket population of stationary states with odd $J$ quantum numbers or $m\neq-J$ are much
      smaller and not shown in the plot. The two-dimensional rotational density contour plot, shown
      in the middle, has been projected onto a sphere, shown on the right hand side, to illustrate
      the orientation of PH$_3$ and the molecular frame relative to the rotation axes. Each
      high-density \emph{island} (red) indicates an axis of rotation and $x,y,z$ refer to the
      molecule-fixed axis system. In the bottom left hand corner, the time evolution of the
      expectation value of the laboratory-fixed effective dipole moment \expectation{\mu_Z} has been
      calculated for the \textit{R} and \textit{L} rotating enantiomers. The exhibited anti-phase
      behavior between the rotating enantiomers confirms chirality.}
   \label{fig:wavepacket}%
\end{figure*}
In \autoref{fig:wavepacket}, the wavepacket population at the end of the centrifuge pulse is plotted
along with the rotational probability density function
$P(\theta,\chi)=\int\dif{V}\dif{\phi}~\psi(t)^{\ast}\psi(t)\sin\theta$, which illustrates the
orientation of the molecule relative to the possible axes of rotation. Here, $\theta,\chi,\phi$
denote the Euler angles, $\dif{V}$ is the volume element associated with the vibrational
coordinates, and $\psi(t)$ is the wavepacket. The dominant contribution of $\ordsim25~\%$ to the
wavepacket is from the $J=42,m=-42$ energy level of the top cluster state at 7686~cm$^{-1}$. In
PH$_3$, clustering occurs above $J_\text{c}\approx35$ and is related to the six islands that emerge
on the rotational density plot seen in \subautoref{fig:wavepacket}{a}, which correspond to clockwise
($\theta<\pi/2$) and anticlockwise ($\theta>\pi/2$) rotation about an axis almost coinciding with
one of the P--H bonds. Application of a strong dc field in the $-Z$ or $+Z$ direction for the
duration of the centrifuge pulse, shown in \subautoref{fig:wavepacket}{b}, produces only clockwise
(\textit{R}) or anticlockwise (\textit{L}) rotating enantiomers, respectively. This is a result of
the brute-force orientation of the permanent dipole moment at the time when population is
transferred into the cluster states. A detailed analysis of the effect of the dc field shows that,
in principle, the field is only necessary during the time when the cluster states are populated,
whereas it is a loss channel at the start of the centrifuge, and it does not affect the dynamics
once the population is within a well-defined angular momentum enantiomer. In order for the
population to be adiabatically transferred from the quasi-free-rotor state to the mixed-parity
`single-enantiomer' cluster state, the dc field splitting must be large in comparison to the
traversal time of the $J$ manifold by the centrifuge. Classically speaking, the dc field has to be
strong enough to favor one enantiomer over the other when the dipole moment is still nearly
perpendicular to the dc field.

This mixed-field enantiomer selectivity allows us to probe the chirality of the \textit{R} and
\textit{L} enantiomers by measuring, for instance, the free-induction
decay~\cite{Patterson:Nature497:475, Yachmenev:PRL117:033001} by a phase-locked microwave
receiver~\cite{Kim:NatPhoton2:733}, see also further details in the supplemental
material~\cite{SM:PRL:PH3:2018}. As expected, the time evolution of the permanent electric dipole
moment expectation value \expectation{\mu_Z} in the laboratory-fixed frame shows anti-phase behavior
between the rotating enantiomers.

If PH$_3$ is excited above $J=42$, the axis of rotation moves closer to the P--H
bond~\cite{Yurchenko:PCCP7:573}. However, our simulations have shown that it is difficult to
efficiently transfer population to higher $J$ states, notably across the range $J\approx40$--$50$.
Instead, wavepacket population gradually spreads across several energy levels and cluster states at
each excitation of $J$, resulting in a spectrally broadened rotational wavepacket. For example, if
the centrifuge pulse used in this work is left to propagate until $t=48.5$~ps, only $1$~\% of
population is transferred to the $J=58,m=-58$ energy level of the top cluster state at
14\,133~cm$^{-1}$. Clustering represents a significant change in the rotational dynamics and as
PH$_3$ rotates faster its structure distorts, leading to the breakdown of $\mathbf{C}_{3v}$(M)
molecular symmetry. The quantum number $k$ is no longer a good quantum number and the $\Delta{k}=0$
selection rule that partially governed transitions up to $J_\text{c}$ becomes weak. The spreading of
wavepacket population can be marginally counteracted with a more tailored centrifuge
pulse~\cite{Owens:JPCL9:4206}, \eg, non-linear angular acceleration and intensity chirping, however,
for the purposes of this study it is enough to show that PH$_3$ is indeed chiral when excited to the
$J=42,m=-42$ cluster state.

The concept of true and false chirality was introduced to clarify situations where motion played an
essential role~\cite{Barron:MolLightScat}. Truly chiral systems possess two distinct enantiomeric
states that can be interconverted by space inversion, but not by time reversal combined with any
proper spatial rotation.
\begin{figure}
   \centering%
   \includegraphics[width=\linewidth]{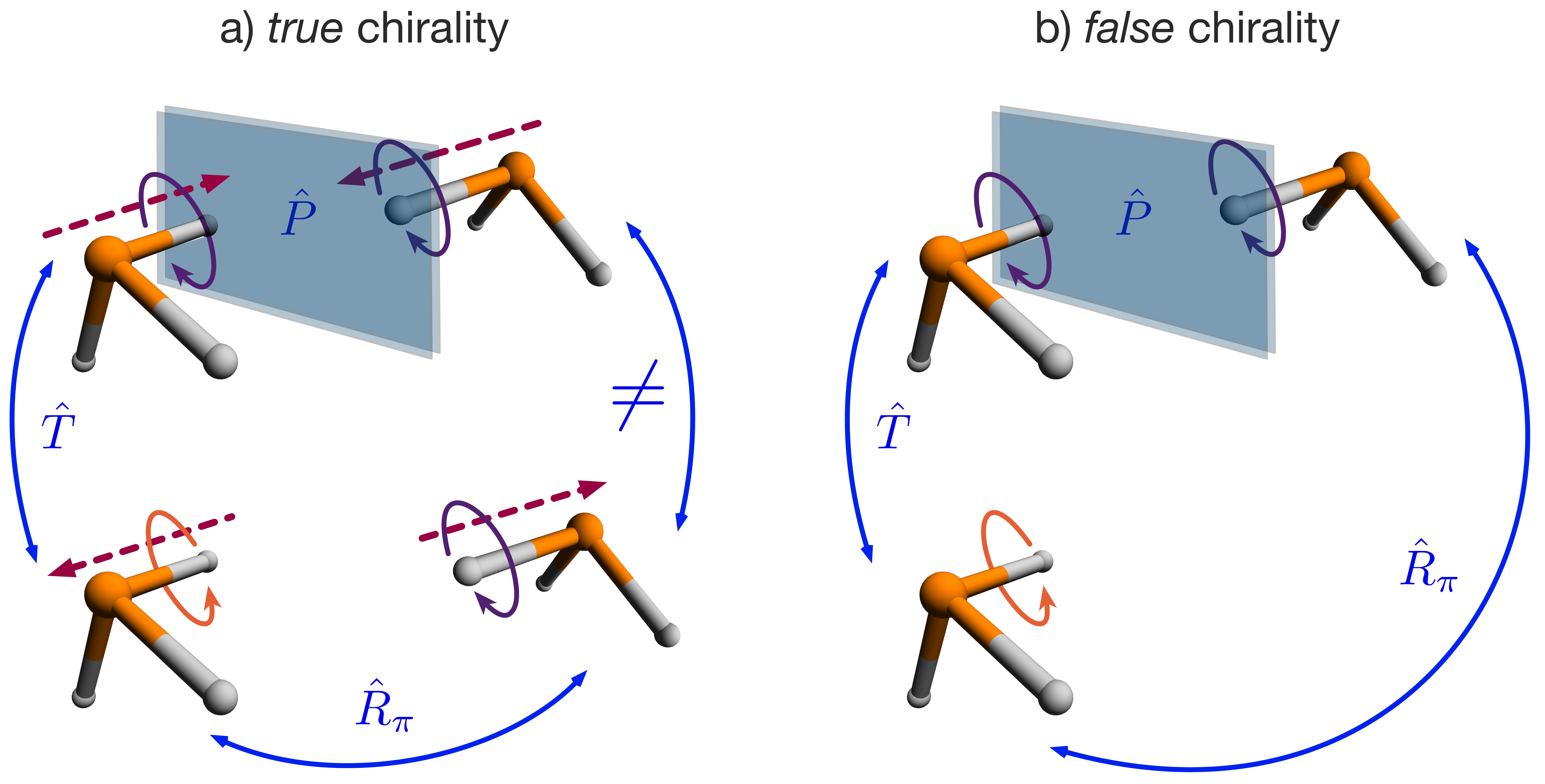}
   \caption{Definition of true and false chirality. (a) If PH$_3$ is translating along the axis of
      rotation (dashed arrow), space inversion ($\hat{P}$) is not equivalent to time reversal
      ($\hat{T}$) combined with any proper spatial rotation ($\hat{R}_{\pi}$) and the system is
      truly chiral. (b) If PH$_3$ is not translating, the system exhibits false chirality as space
      inversion is equivalent to time reversal combined with spatial rotation.}
   \label{fig:true_false}
\end{figure}
As shown in \autoref{fig:true_false}, if PH$_3$ can be forced to translate along the axis of
rotation then the system is truly chiral, \ie, exhibits time-invariant enantiomorphism. If not, the
system possesses false chirality, \ie, time-noninvariant
enantiomorphism. Distinguishing between true and false chirality has implications for topics such as
parity violation~\cite{Barron:CPL123:423} and absolute enantioselection~\cite{Barron:JACS108:5539},
and there is still a great deal to learn about this distinction. For example, it was recently
reported that a falsely chiral influence can generate an enantiomeric excess in a reaction far from
equilibrium~\cite{Micali:NatChem4:201}. In PH$_3$ there is an opportunity to further explore the
differences between true and false chirality by probing the system with and without translation, or
by studying isotopically substituted species such as PHD$_2$, which will possess truly chiral
cluster states even when it is not translating along the axis of rotation.

To establish how, if at all, RIC differs from traditional ``static'' chirality, it will be necessary
to study the interaction of the rotating enantiomers with other chiral entities such as light.
Achieving a deeper understanding should contribute to the development of chirality-based molecular
and material properties, new states of matter, and the potential utilization of RIC in novel
metamaterials or optical devices. It is also of interest to image RIC using sensitive techniques
such as photoelectron circular dichroism (PECD)~\cite{Janssen:PCCP16:856}. It has been shown that
changes in the nuclear geometry caused by vibrational motion can reverse the forward-backward
asymmetry in the photoelectron angular distribution~\cite{Garcia:NatComm4:2132}. One can, therefore,
expect that PECD measurements of RIC may reveal unexpected behavior. We plan to investigate these
topics in future work.

\begin{acknowledgments}
   This work has been supported by the \emph{Deutsche Forschungsgemeinschaft} (DFG) through the
   excellence cluster ``The Hamburg Center for Ultrafast Imaging -- Structure, Dynamics and Control
   of Matter at the Atomic Scale'' (CUI, EXC1074) and the priority program 1840 ``Quantum Dynamics
   in Tailored Intense Fields'' (QUTIF, KU1527/3), by the European Research Council through the
   Consolidator Grant COMOTION (ERC-Küpper-614507), by the Helmholtz Association ``Initiative and
   Networking Fund'', and by the COST action MOLIM (CM1405). A.~O.\ gratefully acknowledges a
   fellowship from the Alexander von Humboldt Foundation.
\end{acknowledgments}
\bibliography{string,cmi,local}
\onecolumngrid
\end{document}